\documentclass[a4paper,fleqn]{cas-dc}

\usepackage[authoryear,longnamesfirst]{natbib}

\usepackage{caption}
\usepackage{tabularray}
\usepackage{subcaption}

\def\tsc#1{\csdef{#1}{\textsc{\lowercase{#1}}\xspace}}
\tsc{WGM}
\tsc{QE}
\tsc{EP}
\tsc{PMS}
\tsc{BEC}
\tsc{DE}

\begin{document}
\let\WriteBookmarks\relax
\def\floatpagepagefraction{1}
\def\textpagefraction{.001}

\shorttitle{Ship Mounted Wave Measurements}

\shortauthors{{\O}lberg et~al.}

\title [mode = title]{Wave Measurements using Open Source Ship Mounted Ultrasonic Altimeter and Motion Correction System during the One Ocean Expedition}     



%
\author[1,2]{Judith Thu {\O}lberg}[orcid=0009-0006-1309-2305]
\author[1]{Patrik Bohlinger}[orcid=0000-0002-7299-3676]
\author[1,2]{{\O}yvind Breivik}[orcid=0000-0002-2900-8458]
\author[3,4]{Kai H. Christensen}[orcid=0000-0002-5775-794X]
\author[1,2]{Birgitte R. Furevik}[orcid=0000-0001-7822-3165]
\author[1]{Lars R. Hole}[orcid=0000-0002-2246-9235]
\cormark[1]
\ead{lrh@met.no}
\author[1]{Gaute Hope}[orcid=0000-0002-5653-1447]
\author[5]{Atle Jensen}[orcid=0000-0001-7376-0850]
\author[6]{Fabian Knoblauch}[orcid=0009-0002-1538-7802]
\author[7]{Ngoc-Thanh Nguyen}[orcid=0000-0003-1576-3345]
\author[3]{Jean Rabault}[orcid=0000-0002-7244-6592]





\affiliation[1]{organization={Norwegian Meteorological Institute},
    addressline={All\'{e}gt. 70}, 
    city={5007 Bergen},
    country={Norway}}
\affiliation[2]{organization={Geophysical Institute, University of Bergen},
    addressline={All\'{e}gt. 70}, 
    city={5007 Bergen},
    country={Norway}}
\affiliation[3]{organization={Norwegian Meteorological Institute},
    addressline={Henrik Mohns plass 1}, 
    city={0371 Oslo},
    country={Norway}}
\affiliation[4]{organization={University of Oslo, Dept. of Geosciences},
    addressline={Postbox 1022}, 
    city={0315 Oslo},
    country={Norway}}
\affiliation[5]{organization={University of Oslo, Dept. of Mathematics},
    addressline={Postbox 1053}, 
    city={0851 Oslo},
    country={Norway}}
\affiliation[6]{organization={Norwegian University of Science and Technology, Department of Civil and Environmental Engineering},
    addressline={Postbox 8900}, 
    city={7491 Trondheim},
    country={Norway}}
\affiliation[7]{organization={Western Norway University of Applied Sciences},
    addressline={Postbox 7030}, 
    city={5020 Bergen},
    country={Norway}}



\cortext[cor1]{Corresponding author}


\begin{abstract}
This study reviews the design and signal processing of ship borne ultrasonic altimeter wave measurements. The system combines a downward facing ultrasonic altimeter to capture the sea surface elevation as a time series, and an inertial measurement unit to compensate for the ship's motion. The methodology is cost-effective, open source, and adaptable to various ships and platforms. The system was installed on the barque Statsraad Lehmkuhl and recorded data continuously during the 20-month One Ocean Expedition. Results from 1-month crossing of the Tropical Atlantic are presented here. The one-dimensional wave spectrum and associated wave parameters are obtained from the sea surface elevation time series. The observed significant wave height agrees well with satellite altimetry and a spectral wave model. The agreement between observations and the spectral wave model is better for the mean wave period than the peak period. We perform Doppler shift corrections to improve wave period estimates by accounting for the speed of the ship relative to the waves. This correction enhances the accuracy of the mean period, but not the peak period. We suggest that the Doppler correction could be improved by complementing the data sources with directional wave measurements from a marine X-band radar.

\end{abstract}




\maketitle

\section{Introduction}

Ocean wave measurements are essential for both scientific and practical purposes. Wave measurements enhance our understanding of the transfer of energy, oxygen, carbon dioxide, and other gases between the atmosphere and the ocean \citep{Deike2021}. Measurements also provide data for calibration and validation to advance satellite remote sensing techniques and numerical wave forecasting \citep{Bidlot2002, Yang2019}. Accurate wave predictions are essential to ensure safe offshore operations and design vessels and offshore structures \citep{Reistad2011}. 

A practical and inexpensive approach for collecting wave measurements is to mount instruments on ships. Ship equipped with sensors can cover great distances of the ocean surface and can navigate to designated study areas, such as the marginal ice zone \citep{Loken2021}. Ship mounted instruments can record autonomously and deployment or retrieval missions are inexpensive. Data can be broadcast directly during research expeditions or conventional shipping routes. Instrumentation may include altimeters operating with radar, laser, or sonic technology \citep{Cifuentes-Lorenzen2013}, wave staffs \citep{Drennan1994}, GPS recording small scale ship motion \citep{Collins2015} and marine X-band radars \citep{Lund2016,Liu2016}. 

The methodology presented here combines a downward facing ultrasonic altimeter and a motion correction device \citep{Christensen2013}. Downward facing altimeters are commonly mounted on offshore platforms to measure the sea surface elevation \citep{Ewans2014, Bohlinger2023, Malila2022}. They are easy to deploy, maintain and recover, as they are not in contact with the water. However, measurements from moving platforms, such as a ship, aircraft, or autonomous surface vehicles, are contaminated by vertical and angular accelerations and the Doppler effect \citep{Amador2022,Sun2005,Pettersson2003}. The motion contamination on the wave measurements is therefore attempted to be compensated.

Previous studies employing ship mounted ultrasonic altimeters found good agreements for integrated wave parameters, compared to in situ observations, wave model, and satellite altimetry \citep{Loken2021,Christensen2013}. A study applying ship mounted laser altimeter reports better accuracy than marine X-band radar for significant wave height but lower accuracy for peak periods \citep{Lund2017}. These studies focused on periods when the ship was stationary to avoid Doppler contamination. \cite{Cifuentes-Lorenzen2013} Doppler corrected the laser altimeter measurements but considered only periods when the vessel is steaming into waves to avoid ambiguities arising when the vessel is steaming with waves. Based on the work of \citep{Collins2016}, we present a method for Doppler correcting measurements when steaming with waves.

This study summarizes the work of \cite{Knoblauch2022,Olberg2023} and presents the wave measurement system mounted on the barque Statsraad Lehmkuhl. The system is produced in-house from open source hardware, firmware, and postprocessing building blocks. The instrumentation, design and signal processing are described in Section \ref{sec:method} and made freely available under an open source license \ref{sec:opem_source_code}. Data acquisition is outlined in Section \ref{sec:data_aquisition}. The results are presented and discussed in Section \ref{sec:results}. Wave parameters are quality controlled in Section \ref{sec:quality_control}, and compared to satellite altimetry and a numerical wave model in Section \ref{sec:data_comparison}. Concluding remarks are given in Section \ref{sec:conclusion}.

\section{Method}\label{sec:method} 

\subsection{Instrumentation}

\begin{figure*}[ht]
	\centering
	\includegraphics[width=1\textwidth]{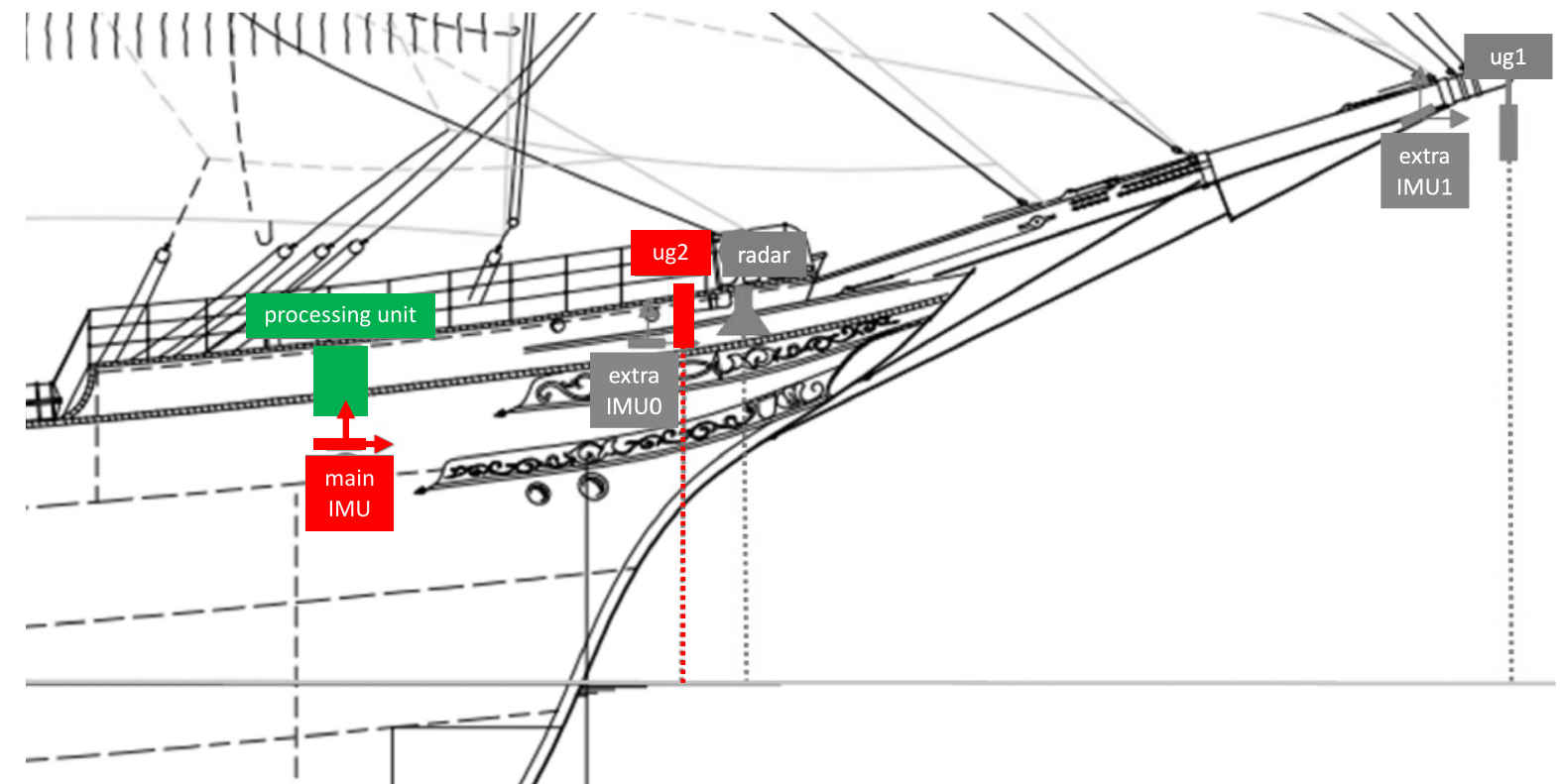}
	\caption{The system components marked on the sketch of Statsraad Lehmkuhl \citep{Knoblauch2022}. Only the colored instruments will be presented here. The extra, redundant instruments not discussed here were installed as a backup in case of individual instruments failure, and as a way to test different configurations to establish how to gather the highest quality data.}
	\label{fig:system_components}
\end{figure*}

The ship mounted system combines altimeter probes to measure the distance from the ship to the ocean surface and inertial measurement units (IMUs) to compensate for the ship's motion. Figure \ref{fig:system_components} illustrates the instruments mounted at the bow of Statsraad Lehmkuhl. Several instruments were installed at different locations on the ship to determine the configuration that collects the highest quality data. The extra instruments also serve as a backup in case of individual instrument failure, which was expected due to the harsh weather conditions encountered during a 20-month-long circumnavigation. 

\begin{figure*}[ht]
	\centering
    \begin{subfigure}{0.445\textwidth}
        \includegraphics[width=1\textwidth]{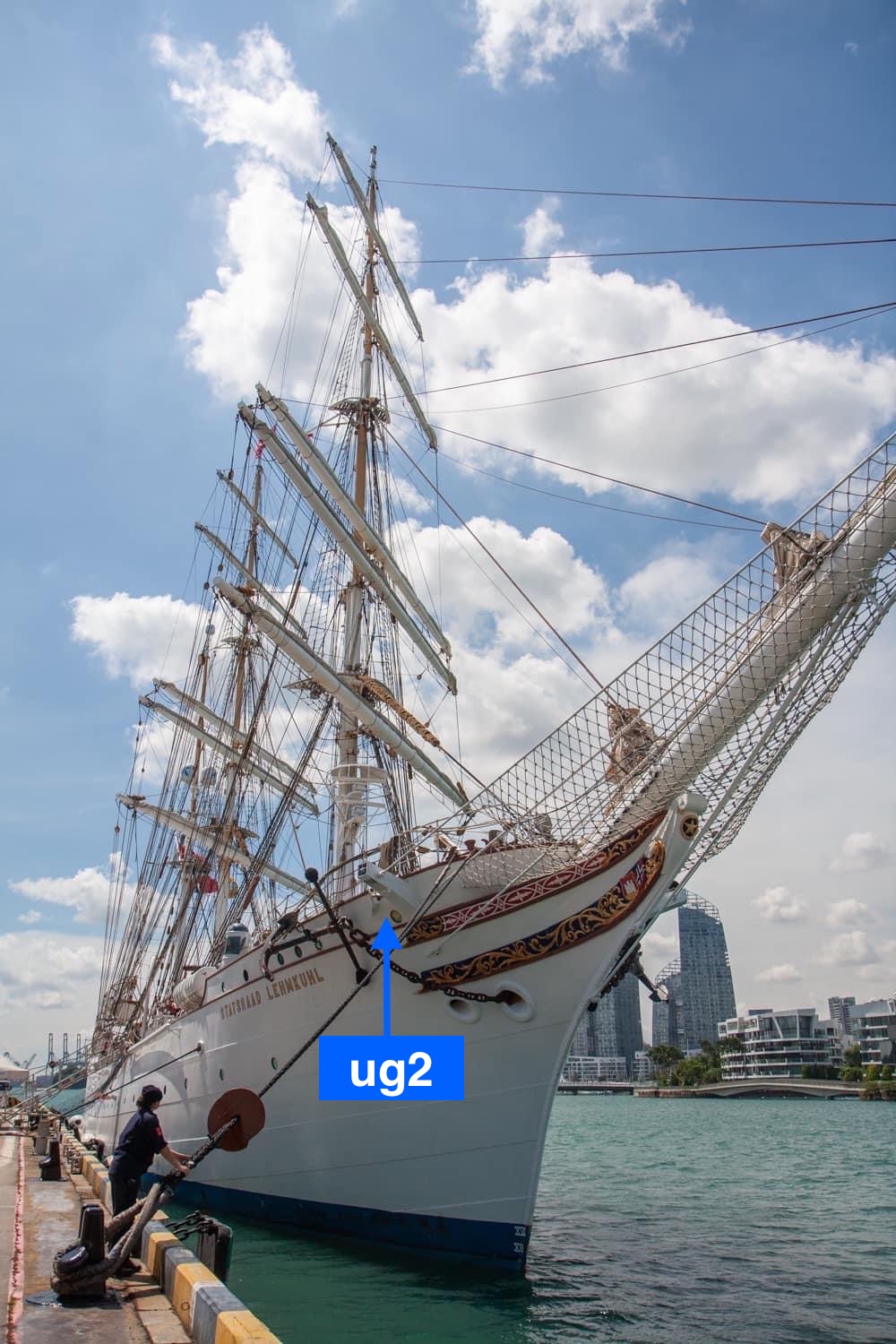}
        \caption{}
	  \label{fig:picture_SL}
    \end{subfigure}
    \begin{subfigure}{0.5\textwidth}
        \includegraphics[width=1\textwidth]{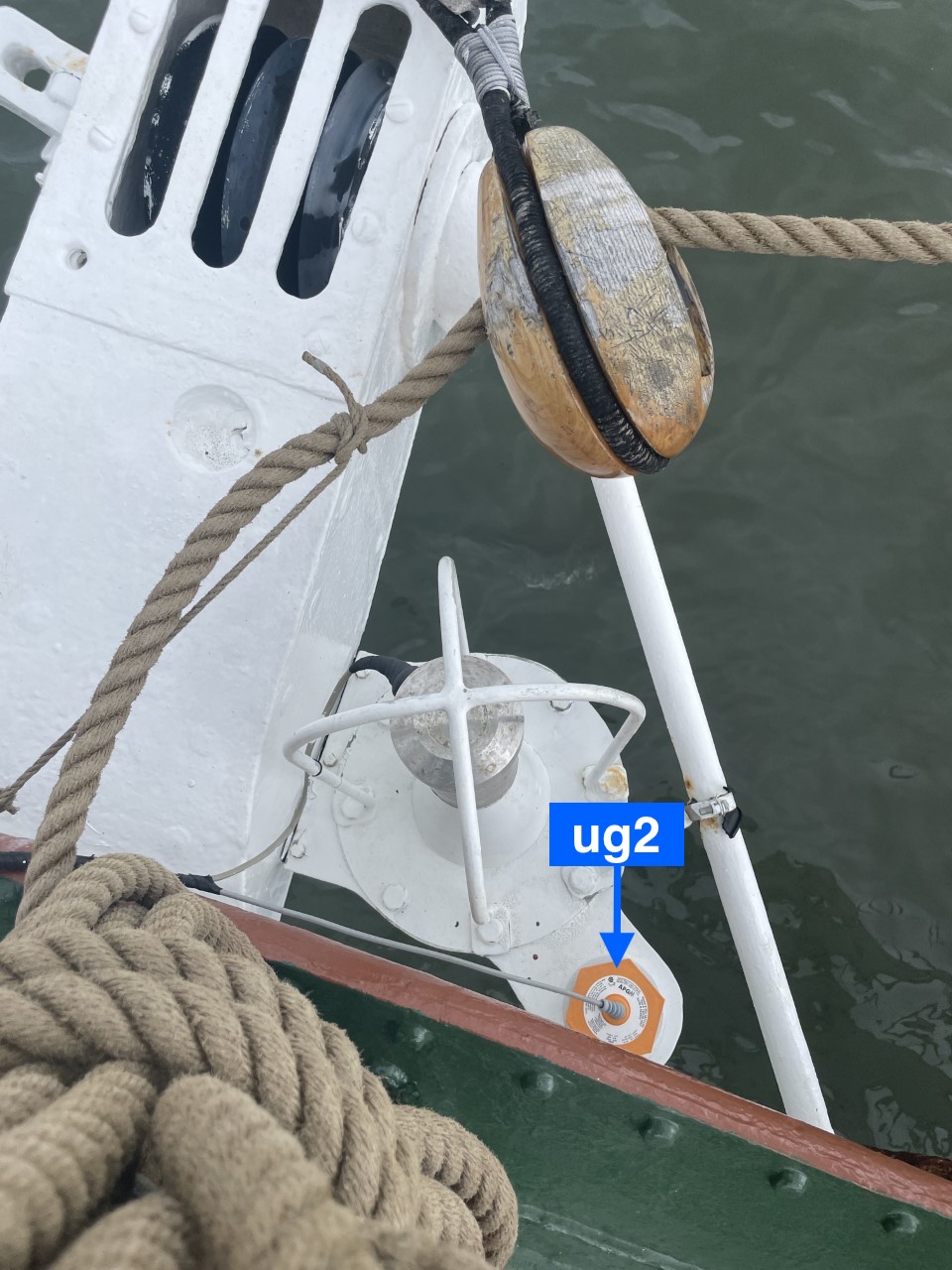}
        \caption{}
	  \label{fig:picture_sensor}
    \end{subfigure}
\caption{Pictures of the position of the ug2 sensor (a) on the starboard bow in (b) a steel frame. Photo: Judith Thu {\O}lberg.}
\label{fig:pictures}
\end{figure*}

The highest accuracy compared to model and satellite data was found for the ultrasonic probe mounted on the starboard side of the bow (ug2) and the main IMU located with the processing unit below deck \citep{Olberg2023}. The two extra IMUs were mounted in waterproof casings exposed at the altimeter probe locations. They were destroyed in harsh storms after respectively three and eight months. Their recordings were subject to more errors than the main IMU \citep{Knoblauch2022}, likely due to vibration contamination at the exposed locations. This could be remediated by enclosing them in heavier and stronger housings giving better protection during storms. The probe operating with microwaves (radar) exhibited a low accuracy, anticipated to be caused by reflections at the ship's metal hull or errors in the set-up of the instrument since the distance fluctuation was inverted compared to the ultrasonic probes \citep{Olberg2023}. The ultrasonic probe mounted at the bowsprit tip (ug1) also contained erroneous measurements due to the extensive motion amplitudes and vibrations encountered at the bowsprit of a sailing ship. 

Therefore, only the colored instruments in Figure \ref{fig:system_components} will be considered in this study. The ultrasonic probe is mounted in a steel frame with a clear path to the ocean surface, as shown in Figure \ref{fig:pictures}. The main IMU is protected from vibrations and physical harm below deck. The horizontal and vertical distance between the main IMU and ug2 are about 4.5 and 1 m, respectively. 

The ultrasonic probe IRU-3433-C60\footnote{\url{https://www.apgsensors.com/wp-content/uploads/2023/04/IRU-manual.pdf}} emits 43 kHz ultrasonic pulses at a 10 Hz sampling rate. It has a distance range of 0.4-15.2 m, and a beam width of 9\textdegree, corresponding to a 0.4-4.89 m footprint diameter at the ocean surface. The distance to the ocean surface is calculated internally from the time delay of the echo, compensated for the speed of sound in air with an internal thermometer. The sensor outputs an analog signal in amperage (4-20mA standard industrial current loop reading).

\begin{figure}[ht]
	\centering
	\includegraphics[width=1\linewidth]{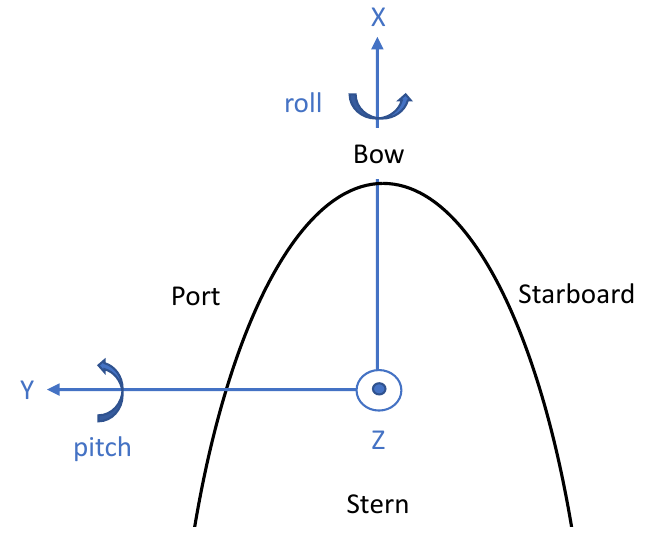}
	\caption{The coordinate system defined for the ship as seen from above. }
	\label{fig:coordinate_system}
\end{figure}

The IMU VN100\footnote{\url{https://www.vectornav.com/products/detail/vn-100}} incorporates an accelerometer, gyroscope, and magnetometer with signal processing capabilities. It samples at 800 Hz and applies internal Kalman- and low-pass filtering \citep{Rabault2020}, providing a digital signal at 10 Hz. The IMU's purpose is to measure the ship's motion. The coordinate system of the ship is defined with the x-axis aligned from stern to bow and the y-axis from starboard to port, illustrated in Figure \ref{fig:coordinate_system}. Roll represents rotation around the x-axis, increasing if the ship tilts from port to starboard. Similarly, pitch represents rotation around the y-axis, increasing if the bow is moved downward during the rotation. The vertical acceleration is positive upwards. The IMUs' coordinate system is defined with the z-axis pointing downwards, mirrored around the x-axis compared to Figure \ref{fig:coordinate_system}. Therefore, the IMUs' vertical acceleration and pitch sign is inverted during postprocessing, while the roll retains its original value. 

\begin{figure}[ht]
	\centering
	\includegraphics[width=1\linewidth]{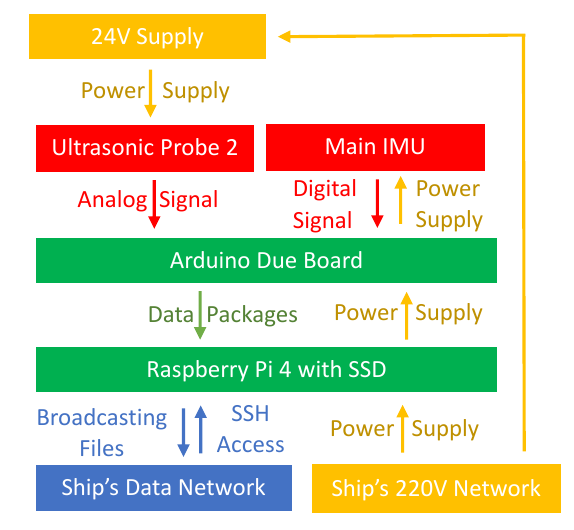}
	\caption{Diagram of the systems' data flow and power supply \citep{Knoblauch2022}.}
	\label{fig:data_flow}
\end{figure}

The processing unit comprises affordable, readily available, and well-documented hardware: an Arduino Due Board\footnote{\url{https://store.arduino.cc/products/arduino-due}} and a Raspberry Pi computer (RPi)\footnote{\url{https://datasheets.raspberrypi.com/rpi4/raspberry-pi-4-product-brief.pdf}}. Figure \ref{fig:data_flow} illustrates the systems data flow and power supply. The RPi runs the processing scripts and saves data as compressed lzma files every 30 minutes. The RPi is protected and fan-cooled by a housing (Argon ONE M.2 case) fitted with an SSD and a real-time clock. It receives power via 5.25 V USB-C (to avoid dropping below 5 V even in the present of resistive loss in cables) from a converter plugged into the ship's 220 V network. The RPi is also linked to the ship's data network via an ethernet cable which provides access via SSH connection and broadcasting data to the ship's internal network and servers. 

The Arduino Due board processes the analog signal from the ultrasonic probe. It performs digital to analog conversion (DAC) of the 4-20 mA current loop signal by using calibrated resistors and measuring their voltage drop. The analog signal is sent as binary data packages to the Raspberry Pi via an actively powered USB hub. The main IMU gets power from the power output pin of the Arduino Due. The ultrasonic probe receives power from a separate 24 V converter plugged into the ship's 220 V network. An electric circuit shield containing a Maxim Integrated MAX3323EEPE is pinned on the Arduino Due to handle voltage shifting together with the resistors for amperage to voltage transformation, and power distribution (schematic circuit diagram is shown in Appendix \ref{sec:circuit_diagram}).  

The total cost of the system comprising the processing unit, ultrasonic probe, and VN100 IMU is about 2500 USD. The price of each individual system component is not presented since it depends on the distributor and individual market situation. However, some general price comparisons will be made since the ship mounted system aims to be cost-effective. The ultrasonic probe is much cheaper than the radar probe\footnote{\url{https://www.apgsensors.com/wp-content/uploads/2023/04/PRL-PRS-manual.pdf}} and has provided better quality data at the same mounting location as the radar during the circumnavigation \citep{Olberg2023}. The price of the system could be cut by using ISM330DHCX-based IMUs\footnote{ \url{https://www.adafruit.com/product/4502}} and custom Kalman filtering and signal processing similar to what is presented by \citet{Rabault2022}. The extra IMUs in the present study were ISM330DHCX-based, but provided more erroneous measurements than the VN100 IMU \citep{Knoblauch2022}. This is likely because the VN100 IMU is better housed and fixed to the ship, not that the instrument is intrinsically better. The ISM330DHCX-based IMUs have proven to give the same data quality as the VN100 IMU at a fraction of the cost when properly installed using robust housing \citep{Rabault2022}. 


\subsection{Signal synchronization and Processing}

To obtain a time series of the surface elevation, measurements from the different sensors, in particular in our case ug2 and the VN100 IMU, are needed on the same time base. The requirement "on the same time base" is not, strictly speaking, exact: it is enough for obtaining usable data that the typical time lag between the measurement acquisition from different sensors is i) predictable, and much smaller than both ii) the typical measurements sample rate and iii) the time scale of the wave dynamics we aim to measure. All measurements, both from IMUs and range sensors, are gathered by the Arduino Due, which runs a bare-metal no-OS program that introduces very low and predictable latency. As a consequence, logging of all measurements is performed consistently by a single hard real-time system without the risk of introducing unexpected delays and time shifts, and the conditions i)-iii) above are met. Indeed, the timing of the range sensors reading is determined by the scheduling of the ADC reading, which is performed and controlled by the Arduino Due bare metal program and allows near-exact signal synchronization (the only time shift introduced comes from the need to actually perform ADC conversions one after the other and to change the ADC channel; in the default Arduino Due ADC configuration, this amounts to around 100 microseconds). The only possible source of time mismatch comes from the IMUs, since their measurement timings are determined intrinsically by each IMU rather than by the program running on the Arduino Due. Therefore, to exactly synchronize time between the range sensors and the IMUs, a common time base is defined as 30 minutes with a time step of 10 Hz, and the ug2 and IMU 30-minute data sets  are thereby linearly interpolated over the common time base.

\begin{figure*}[ht]
	\centering
    \begin{subfigure}{0.45\textwidth}
	   \includegraphics[width=1\linewidth]{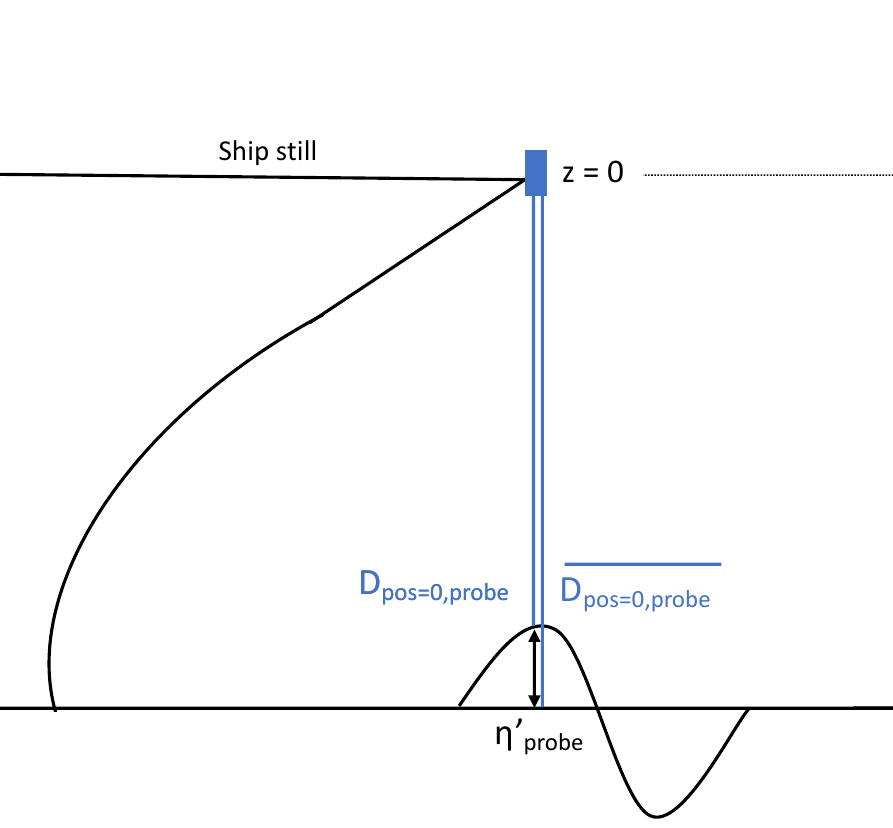}
        \caption{}
        \label{fig:eta_still}
    \end{subfigure}%
    \begin{subfigure}{0.456\textwidth}
	   \includegraphics[width=1\linewidth]{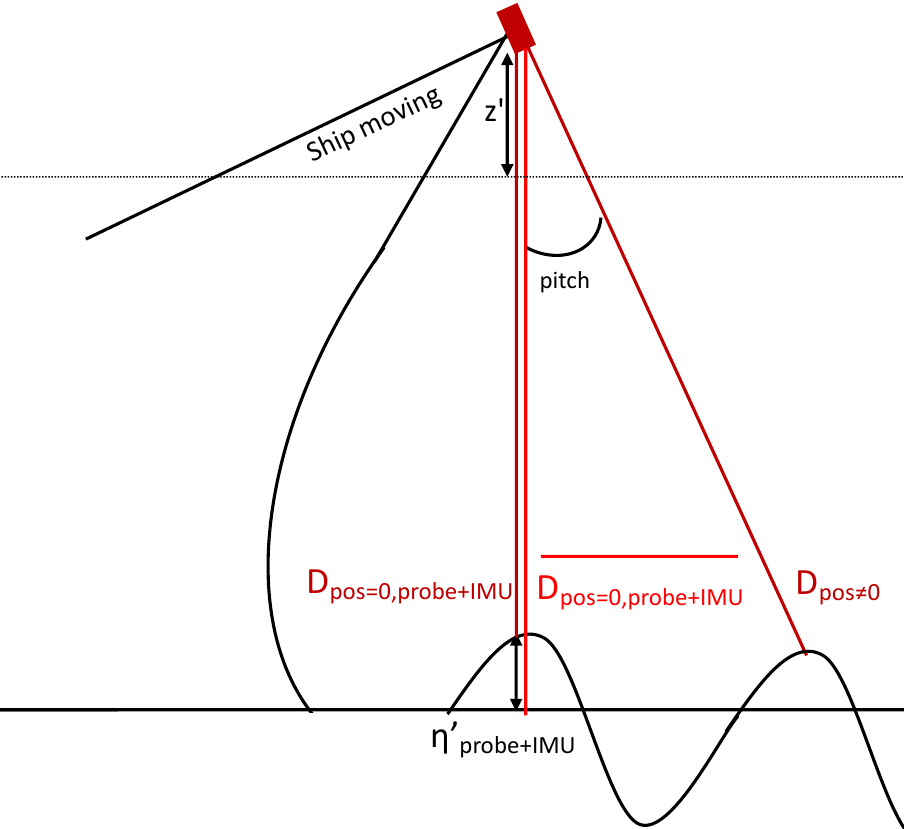}
        \caption{}
        \label{fig:eta_moving}
    \end{subfigure}
\caption{Illustration of the calculation of the water surface fluctuation $\eta$', for a ship (a) lying still in the water and (b) moving due to waves.}
\label{fig:eta_method}
\end{figure*}

\begin{figure*}[h!]
	\centering
	\includegraphics[width=1\textwidth]{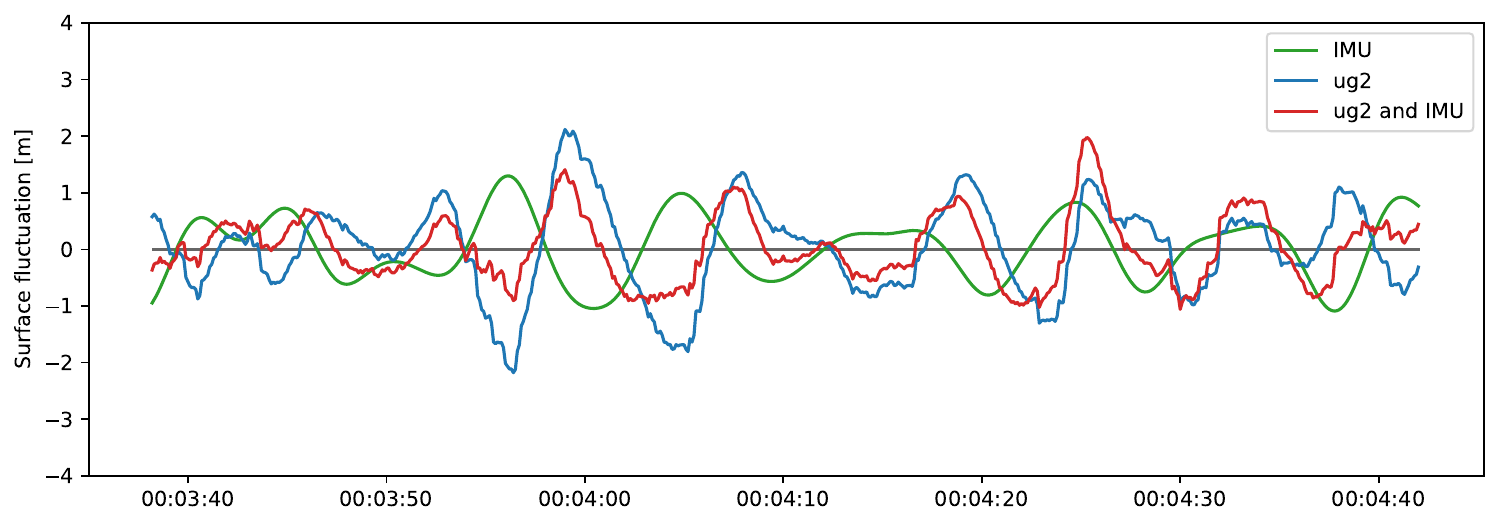}
	\caption{Surface fluctuation over 1-minute the 22.10.21 measured solely by IMU, ug2, and the combination of both. }
	\label{fig:eta_example}
\end{figure*}

The surface fluctuation $\eta'$ would be determined solely by the ultrasonic distance measurement if the ship laid perfectly still in the water, as illustrated in Figure \ref{fig:eta_still}. The marker ' refers to a fluctuation, not a derivative. The surface fluctuation would then be the distance subtracted from the mean distance of each 30-minute file. With this sign convention, a positive fluctuation resembles a crest, and a negative resembles a trough;
\begin{equation}\label{eq:eta_probe}
    \eta'_{_{\textrm{probe}}} = \overline{\footnotesize{D}_{_{\textrm{pos=0,probe}}}} - \footnotesize{D}_{_{{\textrm{pos=0,probe}}}}. 
\end{equation}

However, the ship moves under the influence of waves, which modulates the vertical position and inclination of the ultrasonic probe. Pitch is the inclination in the x-z-plane, while roll is the inclination in the y-z-plane. The trigonometric relations of the orthogonal triangle in Figure \ref{fig:eta_moving} give the distance compensated for the ship's motion similar to Equation 1 in \cite{Christensen2013}:
\begin{equation}\label{eq:eta_probe_imu}
    \footnotesize{D}_{_{\textrm{pos=0,probe+IMU}}} = \footnotesize{D}_{_{\textrm{pos} \neq 0}} \cos(\textrm{\footnotesize{pitch}})\cos(\textrm{\footnotesize{roll}}) - \textrm{z'}.
\end{equation}

Only the surface elevation fluctuation is retained:
\begin{equation}\label{eq:eta_probe_imu_fluc}
    \eta'_{_{\textrm{probe+IMU}}} = \overline{\footnotesize{D}_{_{\textrm{pos=0,probe+IMU}}}} - \footnotesize{D}_{_{\textrm{pos=0,probe+IMU}}}.
\end{equation}

The surface fluctuation can also be determined from only the vertical position of the ship by ignoring the ultrasonic distance measurements of Equation \ref{eq:eta_probe_imu} (D=0):
\begin{equation}\label{eq:eta_imu}
    \eta'_{_{\textrm{IMU}}} = \textrm{z'}.
\end{equation}

The vertical position is obtained by double integrating the IMUs' vertical acceleration. Firstly the gravitational acceleration is subtracted from the vertical acceleration:
\begin{equation}\label{eq:vertical_acceleration}
   \ddot{\textrm{z}}(t)' = \ddot{\textrm{z}}(t) - g.
\end{equation}

The double integration is performed with Fourier transform following:
\begin{equation}\label{eq:vertical_position}
    \textrm{z}(t)' = \mathcal{F}^{-1}\left[\frac{-1}{\omega^2}\mathcal{F}[\ddot{\textrm{z}}(t)'](\omega)\right](t).
\end{equation}

Here, $\omega = 2\pi f$ is the angular frequency \citep{Holthuijsen20073}. The transform is performed by fast Fourier transform for each 30-minute file. Only frequencies inside the range $f_{min} = 0.05$ Hz and $f_{max} = 0.5$ Hz are retained, corresponding to frequencies relevant to ocean waves. The first and last three minutes of each data file are excluded due to potential errors related to the transform.

Figure \ref{fig:eta_example} illustrates an example of the surface fluctuation measured solely by IMU (Equation \ref{eq:eta_imu}), ug2 (Equation \ref{eq:eta_probe}), and the combination of both (Equation \ref{eq:eta_probe_imu_fluc}). \cite{Olberg2023} found, as expected, that the combination of water elevation from ug2 and platform motion correction from the IMU provides the highest accuracy, and is therefore presented in this paper.

\subsection{Wave Spectrum and Wave Parameters}

The one-dimensional variance density spectrum E(f) is determined from the surface fluctuation (Equation \ref{eq:eta_probe_imu_fluc}) by fast Fourier transform with the Welch method \citep{Welch1967}. Each 30-minute file is split into segments of 180 s, corresponding to 1800 sampling points at a 10 Hz sampling rate. Overlap between segments is set to 90 \%. Each segment is multiplied by a Hann window to reduce spectral leakage. The cutoff frequencies are set to $f_{min} = 0.05$ Hz and $f_{max} = 0.5$ Hz.

The wave parameters are calculated from the moments of the variance density spectrum \citep{Holthuijsen20074}:
\begin{equation}\label{eq:mn}
    m_{_n} = \int_{f_{min}}^{f_{max}}f^nE(f)df.
\end{equation}

The integrated significant wave height is determined by the zeroth-order moment of the variance density spectrum $m_{_0}$ \citep{Holthuijsen20074}:
\begin{equation}\label{eq:Hs}
    H_s = 4\sqrt{m_{_0}}.
\end{equation}

The first moment mean period is calculated from the first-order moment \citep{Holthuijsen20074}:
\begin{equation}\label{eq:Tm01}
    T_{{m}_{01}} = \frac{m_{_0}}{m_{_1}}.
\end{equation}

The second moment mean period is calculated to emphasise the high frequency part of the spectrum \citep{Holthuijsen20074}:
\begin{equation}\label{eq:Tm02}
    T_{{m}_{02}} = \sqrt{\frac{m_{_0}}{m_{_2}}}.
\end{equation}

The peak period is defined as the reciprocal of the peak frequency of the variance density spectrum \citep{Holthuijsen20074}, and represents the period of the dominant sea state:
\begin{equation}\label{eq:Tp}
    T_p = \frac{1}{f(max(E(f)))}.
\end{equation}

\subsection{Doppler Effect Correction}

Data recorded while the ship travels contains a Doppler-shifted frequency since the waves are moving relative to the ship. Note that the Doppler shift affects the observed wave frequency, and therefore the peak period (Equation \ref{eq:Tp}), but has no effect on the significant wave height, as it does not affect the moment $m_0$ (Equation \ref{eq:Hs}). The derivation of Doppler corrections for wave frequency spectra follows the work of \cite{Collins2016}. Here, the recorded frequency is called the observer's frequency  $f_o$, while the frequency of the incoming waves is called the sender's frequency  $f_s$. The sender's frequency $f_s$ will be derived from the Doppler relation:
\begin{equation}\label{eq:fo}
    f_o = f_s\left(\frac{c+\textrm{v}_o}{c+\textrm{v}_s}\right).
\end{equation}

\begin{figure}[ht]
	\centering
	\includegraphics[width=1\linewidth]{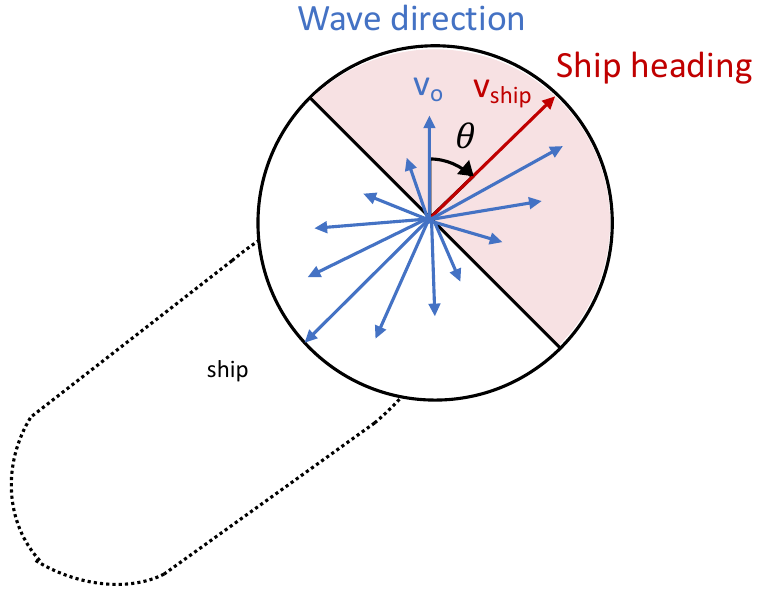}
	\caption{Illustration of the relation between the wave direction, ship heading, and $\theta$.}
	\label{fig:doppler_directions}
\end{figure}

Here, v$_o$ is the observer's velocity in the wave direction. Figure \ref{fig:doppler_directions} illustrates that v$_o$ is the ship's velocity v$_{ship}$ scaled by the angle $\theta$ between the wave direction (coming-from) and the ship heading (going-to).:
\begin{equation}\label{eq:vo}
    \textrm{v}_o = \textrm{v}_{_{\textrm{ship}}} \cdot \cos(\theta).
\end{equation}

The observer's velocity is largest for wave directions aligned with the ship heading (towards or away) and absent for wave directions perpendicular to the ship heading. The velocity is negative in the red-colored area (vessel steaming with waves) and positive in the white space (vessel steaming into waves). The frequencies recorded while steaming with waves are too low and must be shifted to higher values. Similarly, frequencies recorded while steaming into waves are too high and must be corrected to lower frequencies. 

The velocity of the sender v$_s$ is zero because the traveling waves are not caused by a distinct sender. The signal propagating velocity $c$ is the phase speed of surface gravity waves in deep water \citep{Holthuijsen20075}:
\begin{equation}\label{eq:c}
    c = \frac{g}{\omega_s} = \frac{g}{2\pi f_s}.
\end{equation}

Inserting Equation \ref{eq:c} to Equation \ref{eq:fo} gives a second-order equation for the sender's frequency:
\begin{equation}\label{eq:fs2}
    f_s^2 + \frac{g}{2\pi  \textrm{v}_o}f_s - \frac{g}{2\pi \textrm{v}_o}f_o = 0.
\end{equation}

Two solutions can be obtained from Equation \ref{eq:fs2}:
\begin{equation}\label{eq:fs}
    f_{s+,s-} =  \frac{-g \pm \sqrt{g^2+8\pi g\textrm{v}_o f_o}}{4\pi \textrm{v}_o}.
\end{equation}

When the vessel is steaming into waves, the solution is the negative root, and $f_o$ always maps to a lower value of $f_s$ \citep{Collins2016}: 

\begin{equation}\label{eq:fs_neg}
    f_{s-} =  \frac{-g - \sqrt{g^2+8\pi g\textrm{v}_o f_o}}{4\pi \textrm{v}_o}.
\end{equation}

Ambiguities arise when the vessel is steaming with waves because contributions to the observed frequency $f_o$ may originate from three wave components $f_s$ (Figure 1a \cite{Collins2016}). The wave components are long waves propagating faster than twice the ship's speed, short waves being overtaken by the ship, and intermediate waves in between  \citep{Collins2016}. To avoid ambiguities, only the fast waves are Doppler shifted $c_{\textrm{critical}} > \lvert 4 \textrm{v}_o \rvert$ corresponding to low frequencies $f_{\textrm{critical}} < \lvert \frac{g}{8\pi\textrm{v}_o} \rvert $. Frequencies higher than the critical frequency are neglected in the Doppler correction. The negative root (Equation \ref{eq:fs_neg}) is adequate for the fast waves because the square root remains positive, and $f_o$ always maps to a higher value of $f_s$. 

The Doppler shifted periods (Equation \ref{eq:Tm01}, \ref{eq:Tm02}, \ref{eq:Tp}) are thereby calculated from the Doppler shifted frequency spectrum within the range $f_{s_{min}} = 0.05$ Hz and $f_{s_{max}} = 0.5$ Hz.

\begin{figure*}[ht]
	\centering
    \begin{subfigure}{0.49\textwidth}
	   \includegraphics[width=1\linewidth]{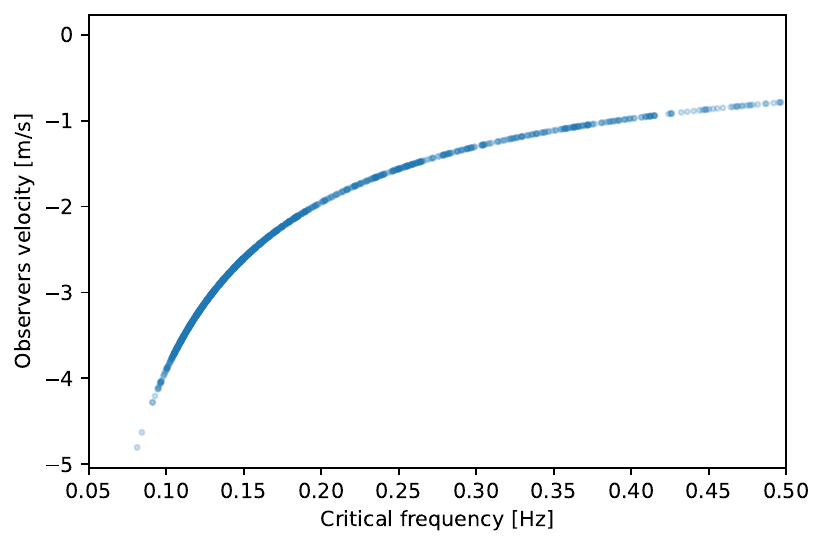}
        \caption{}
        \label{fig:frequency_velocity}
    \end{subfigure}
    \begin{subfigure}{0.49\textwidth}
	   \includegraphics[width=1\linewidth]{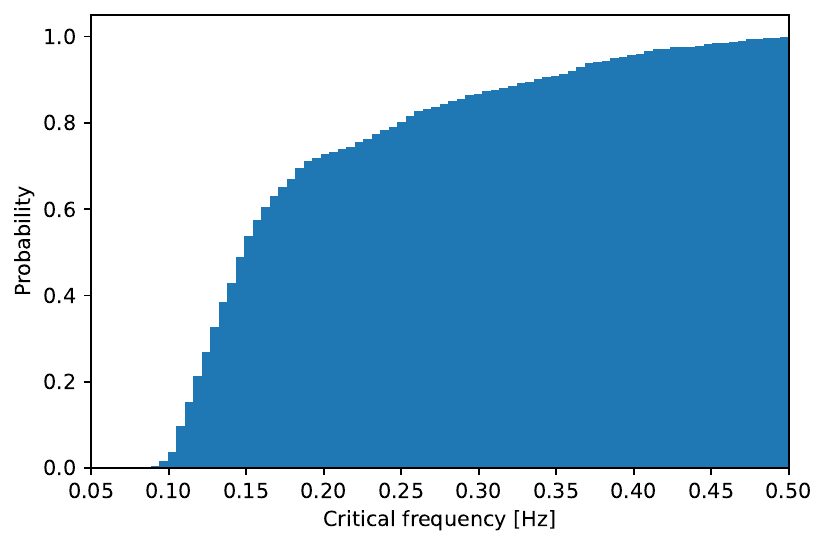}
        \caption{}
        \label{fig:frequency_cumulative}
    \end{subfigure}
\caption{The critical frequency for the Doppler shifted spectrum displayed (a) dependent on the observers velocity and (b) as a cumulative distribution function.}
\label{fig:frequency_critical}
\end{figure*}

Figure \ref{fig:frequency_velocity} illustrates the dependence of the critical frequency as a function of the observers' velocity. At high velocity, the critical frequency is low, and few frequencies are retained in the Doppler shifted spectrum. At low velocity, all frequencies are retained until the cutoff at 0.5 Hz. Figure \ref{fig:frequency_cumulative} displays the cumulative distribution function of the critical frequency. The frequencies below 0.08 Hz are always retained, but already at 0.15 Hz, 50 \% of the Doppler shifted spectra are cut. Steaming with waves therefore results in a reduced spectral range. 

\section{Data Aquisition}\label{sec:data_aquisition}

\begin{figure*}[ht]
	\centering
	\includegraphics[width=1\textwidth]{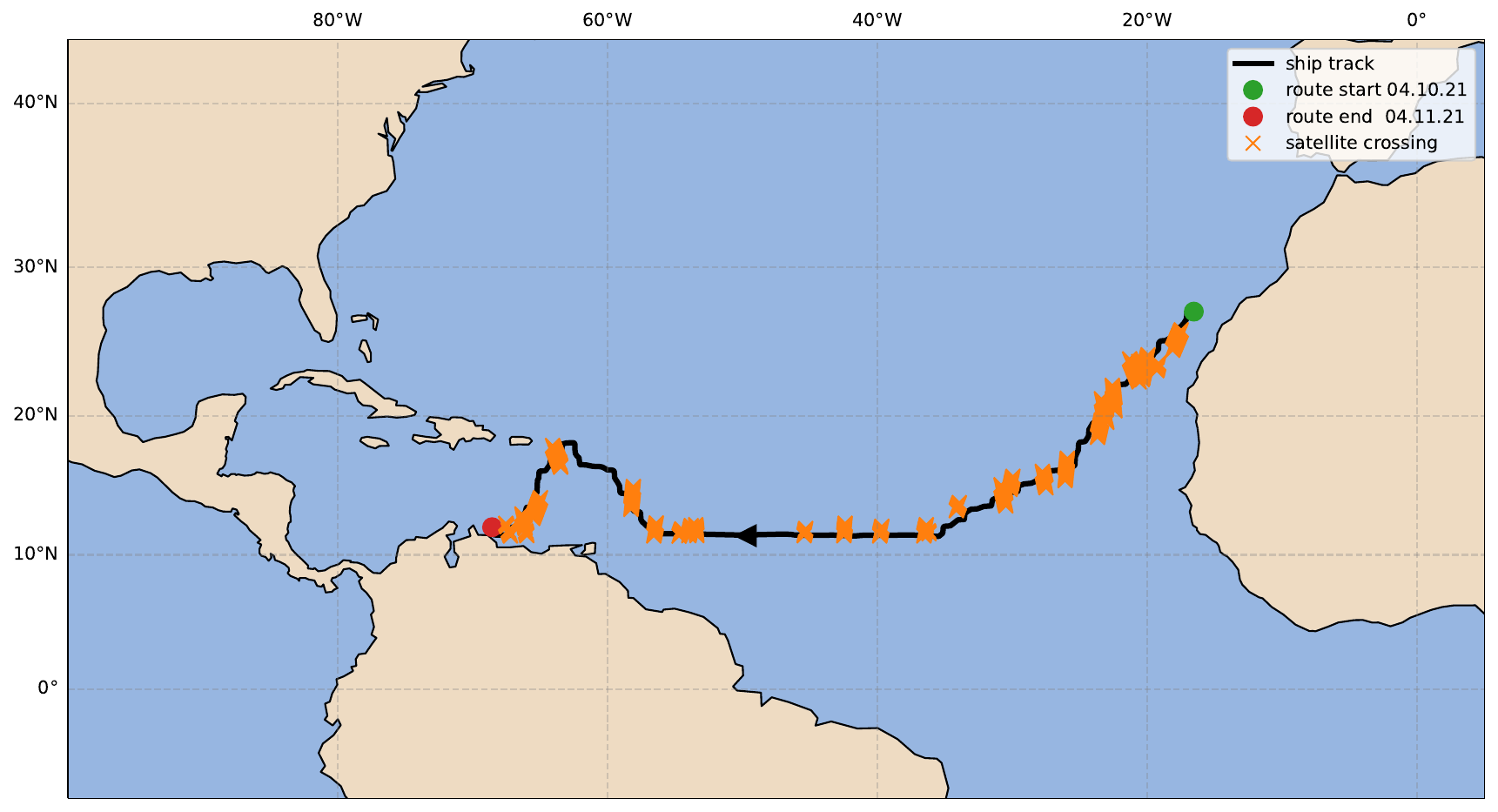}
	\caption{The ship track of Statsraad Lehmkuhl from Las Palmas de Gran Canaria 04.10.21 to Wilhelmstad, Curacao 04.11.21. Orange markers are satellite track crossings measuring wave statistics which are used for data comparison.}
	\label{fig:ship_track}
\end{figure*}

Data for this study were collected during the One Ocean Expedition\footnote{\url{https://oneoceanexpedition.com}}. The 98-meter-long Norwegian barque, Statsraad Lehmkuhl, circumnavigated the globe from Arendal in August 2021 to Bergen in April 2023. The ship was equipped with various scientific equipment, which aims to gather knowledge about the ocean \citep{Huse2023}. All research data are available open source\footnote{\url{http://metadata.nmdc.no/metadata-api/landingpage/16316eb3cea666a1871679a8b78568e1}}, including the data used in this study regarding wave measurements, wind, ship track, speed, and heading. The data presented here were recorded from Las Palmas de Gran Canaria, 04.10.21 to Wilhelmstad, Curacao, 04.11.21, illustrated in Figure \ref{fig:ship_track}. Over this period, the system was recently mounted and functioning optimally, and it's one of the longest continuous wave records. 

The recorded wave parameters are compared to the third-generation wave model ECWAM which is operated by the European Centre for Medium-Range Weather Forecast \citep{ECMWF2020}. ECWAM has global coverage with 14 km spatial- and 1-hour temporal resolution. The model provides a 10-day forecast twice a day. The shortest lead time available (best estimate) is chosen. The wave parameters applied for comparison are the significant wave height of combined wind waves and swell, mean wave period based on first moment, the second moment wave period, peak wave period and mean wave direction (parameter ID 140229, 140220, 140221, 140231, 140230\footnote{\url{https://codes.ecmwf.int/grib/param-db/?filter=grib2}}). ECWAM's mean wave direction is utilized for the Doppler effect calculation because currently the ship mounted cannot measure directional properties. The wave direction is defined as the direction the waves are coming from. 

To complement model data, satellite altimeter measurements are used to validate the ship borne wave measurements. Significant wave height from several satellite missions is accessed from Copernicus Marine Environment Monitoring Service\footnote{\url{https://data.marine.copernicus.eu/product/WAVE_GLO_PHY_SWH_L3_NRT_014_001/description}}. The satellite missions: CryoSat-2, Sentinel-3A and -3B, HaiYang-2B, CFOSAT, and Sentinel-6A, are considered here. These are not used for assimilation in ECWAM and can therefore serve as independent quality control (Abdalla S., personal communication, January 2023). The satellite data are level-3, which is quality-controlled, inter-calibrated, and filtered for noise \citep{Taburet2022}. The filtered product has a spatial and temporal resolution of 7km and 1Hz, respectively.

The model and satellite data are collected along the ship track using the open source software package Wavy\footnote{\url{https://github.com/bohlinger/wavy}}. The Wavy software collects data recorded within a defined time window and distance limit \citep{Bohlinger2019}, chosen as 15 minutes and 100km, respectively. For model data, only the nearest grid cell within the distance limit is extracted. This results in two time series, one for the wave model and one for ship measurements, with an equal number of values. For satellite data, all values within the distance limit are gathered. The value along a local satellite track segment closest in distance from the ship track is chosen, providing two time series with an equal number of satellite values and ship measurements. The Wavy software also calculates validation statistics, including the root mean squared deviation (RMSD), mean deviation (MD), correlation coefficient (R), and the number of observations collocated (N) (see Appendix \ref{sec:validation_statistics}). The Wavy documentation describes an example of collocation with the ship track\footnote{\url{https://wavyopen.readthedocs.io/en/latest/vietnam22.html}}.

\section{Results and Discussion}\label{sec:results}

In the route studied here (Figure \ref{fig:ship_track}), Statsraad Lehmkuhl was sailing with an average wind speed of 6.3 m/s and a maximum of 13.6 m/s. The average ship speed was 2.7 m/s, and the maximum speed was 5.3 m/s without using the motor. The wave direction was mainly toward the southwest. Since the ship was sailing west, it was steaming with the waves. Waves were entering the starboard side of the stern, and since the ug2 sensor is located on the starboard side, shadowing effects by the ship's hull are assumed to be avoided.   

The ultrasonic sensor measured a mean distance of 7.0 m and a 1.2-15.4 m range. This indicates that the ultrasonic sensor was never submerged in the water. The maximum specified sensor distance of 15.2 m was exceeded at 1.5 \% of the measurements, which is a low value compared to previous studies \citep{Christensen2013,Loken2021}. The ultrasonic sensor's mounting and range suit the relatively calm sea conditions presented here. 

\subsection{Quality Control}\label{sec:quality_control}
The collected data were automatically quality controlled by our anomaly detection algorithm called AdapAD~\citep{nguyen2023icse}. AdapAD is designed to work on oceanographic data and aim to detect measurements that are too deviated from the majority of data instances. The anomalous measurements are detected by the AdapAD algorithm and then manually inspected by ourselves.

SalaciaML~\citep{mieruch2021salaciaml} and CoTeDe~\citep{castelao2020framework} are existing software solutions for automatic data quality which also make use of anomaly detection. However, they are not suitable for the present study. SalaciaML, a deep learning classification model, requires an extensive amount of quality-controlled data during development. The sailing route of Statsraad Lehmkuhl is unique so no similar data is readily available not to mention the data must be quality-controlled. CoTeDe requires several pre-defined thresholds to determine normality or anomaly. In contrast, AdapAD does not need to be trained in advance. The performance of the AdapAD algorithm was evaluated on our quality-controlled data. It yields detection precision of at least 87\% and higher than 38 state-of-the-art anomaly detection algorithms.

The AdapAD algorithm is applied to seven wave properties collected by the ship mounted system. Table~\ref{table:data_contaminiation} list the data contamination rate of the wave properties. The data contamination rate is highest for the peak wave period, but makes up less than 1\% for all wave parameters. As the data contamination rate is low, the data is adequate for other data-driven applications ~\citep{schmidl2022anomaly,wu2021current}. The quality-controlled data is accessible in our replication package \ref{sec:opem_source_code}. 


\begin{table*}[width=1\textwidth,pos=ht]
\caption{Data contamination rate of wave properties.}
\label{table:data_contaminiation}
\begin{tabular*}{\tblwidth}{@{} LR@{} }
\toprule
\textbf{Wave property}                                      & \textbf{Data contamination rate} \\
\midrule
Significant wave height (Hs)                                & 1/1440 (0.07\%)         \\
Original first moment mean wave period Tm01 (Tm01)          & 3/1440 (0.21\%)         \\
Doppler shifted first moment mean wave period (Tm01\_ds)    & 7/1440 (0.49\%)         \\
Original second moment mean wave period (Tm02)              & 4/1440 (0.28\%)         \\
Doppler shifted second moment mean wave period  (Tm02\_ds)  & 3/1440 (0.21\%)         \\
Original peak wave period (Tp)                              & 14/1440 (0.97\%)        \\
Doppler shifted peak wave period  (Tp\_ds)                  & 2/1440 (0.14\%)         \\
\bottomrule
\end{tabular*}
\end{table*}

\subsection{Data Comparison}\label{sec:data_comparison}

\begin{figure*}[ht]
\centering
    \begin{subfigure}{0.45\textwidth}
        \includegraphics[width=1\textwidth]{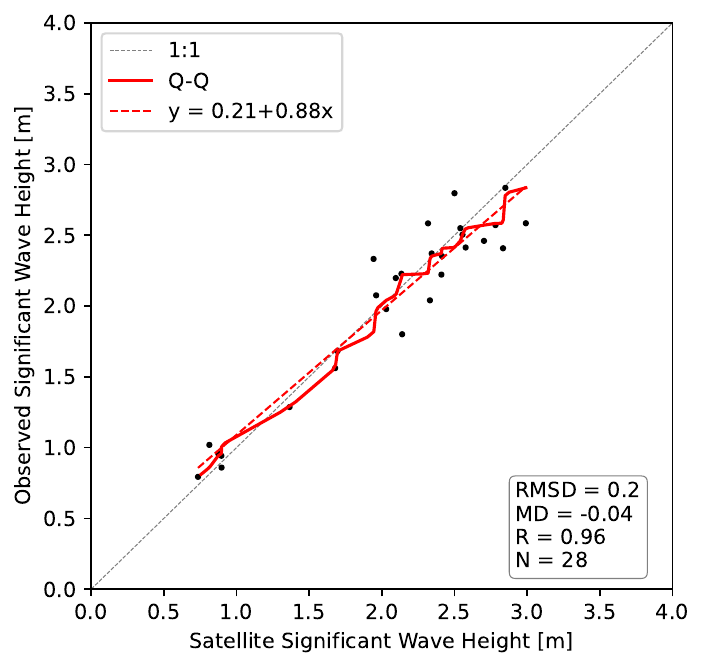}
        \caption{}
	  \label{fig:scatter_hs_sat}
    \end{subfigure}
    \begin{subfigure}{0.45\textwidth}
        \includegraphics[width=1\textwidth]{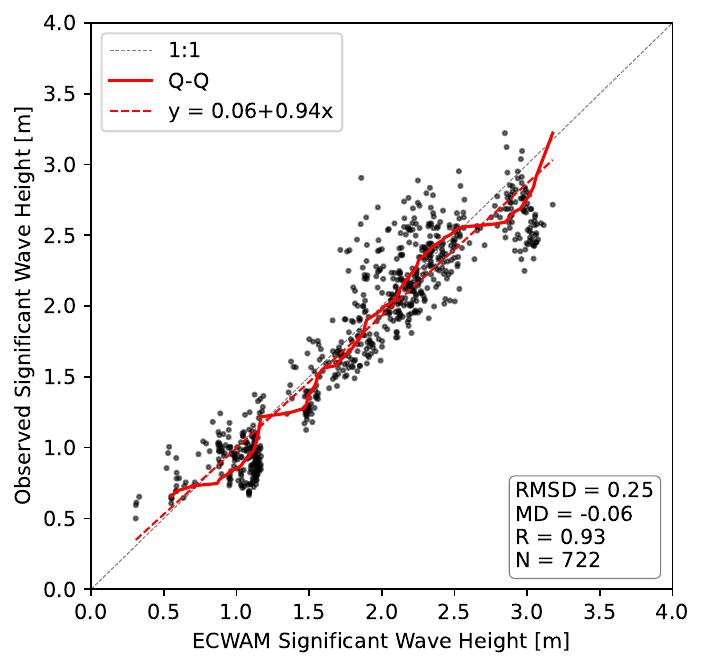}
        \caption{}
	  \label{fig:scatter_hs}
    \end{subfigure}
\caption{Observed significant wave height compared to (a) satellite and (b) ECWAM. The solid red line is the quantile-quantile fit (Q-Q), while the red dotted line is the linear regression (y). The statistics are described in Appendix \ref{sec:validation_statistics}.}
\label{fig:scatter_hight}
\end{figure*}

On average, the observed significant wave height was 1.8 m, with a maximum value of 3.3 m. The significant wave height observations fit well with the 28 satellite observations available alongside the ship track (Figure \ref{fig:scatter_hs_sat}). The root mean squared deviation is 0.20 m and the correlation coefficient is 0.96. The observations also agree well with the wave model data (Figure \ref{fig:scatter_hs}) for the 722 hourly values. The root mean squared deviation is 0.25 m and the correlation coefficient is 0.93.

\begin{figure*}[ht]
	\centering   
    \begin{subfigure}{0.45\textwidth}
        \includegraphics[width=1\textwidth]{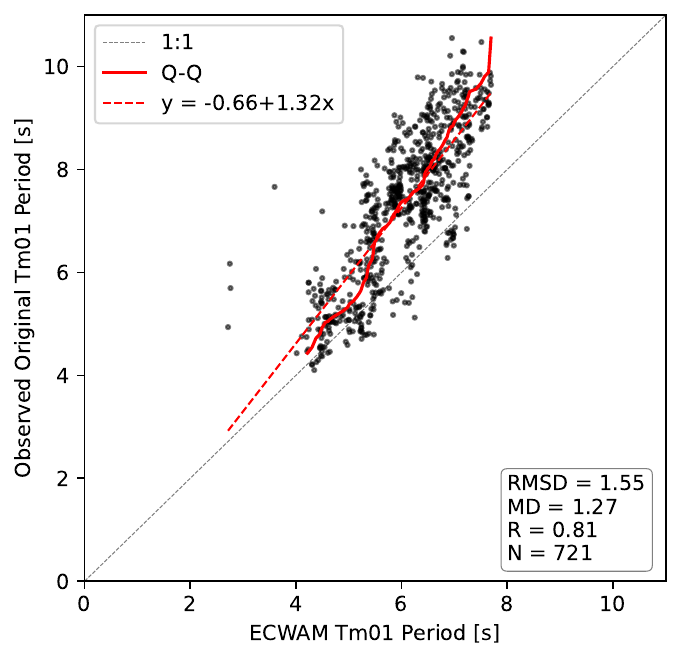}
        \caption{}
	  \label{fig:scatter_tf}
    \end{subfigure}
    \begin{subfigure}{0.45\textwidth}
        \includegraphics[width=1\textwidth]{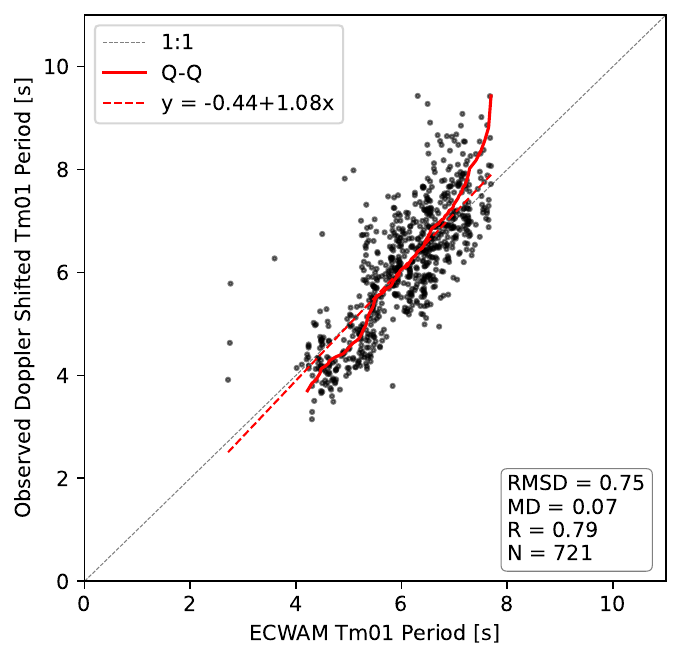}
        \caption{}
	  \label{fig:scatter_tf_ds}
    \end{subfigure} 
    \begin{subfigure}{0.45\textwidth}
        \includegraphics[width=1\textwidth]{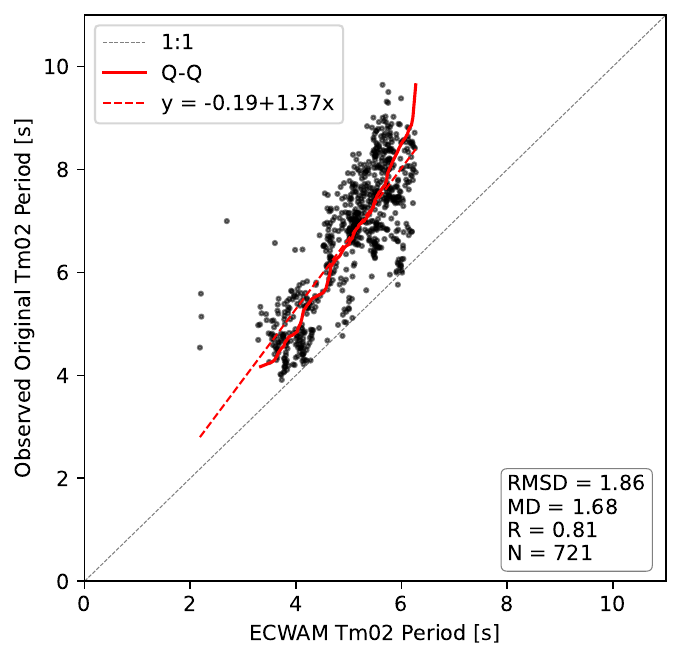}
        \caption{}
	  \label{fig:scatter_tzc}
    \end{subfigure}
    \begin{subfigure}{0.45\textwidth}
        \includegraphics[width=1\textwidth]{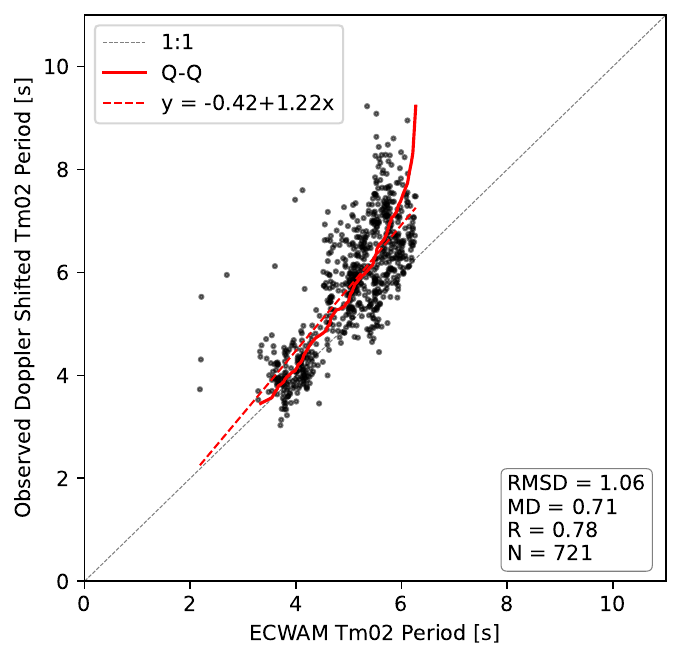}
        \caption{}
	  \label{fig:scatter_tzc_ds}
    \end{subfigure}
\caption{Observed (a) original Tm01 period, (b) Doppler shifted Tm01 period, (c) original Tm02 period and (d) Doppler shifted Tm02 period, compared to ECWAM's respective periods. The solid red line is the quantile-quantile fit (Q-Q), while the red dotted line is the linear regression (y). The statistics are described in Appendix \ref{sec:validation_statistics}.}
\label{fig:scatter_mean_period}
\end{figure*}

The observed first moment mean period Tm01 has a mean value of 7.3 s and a maximum value of 10.6 s. The second moment mean period Tm02 has a lower mean value of 6.7 s and a maximum value of 9.6 s since it emphasizes the high frequencies. Both mean periods agree well with the wave model (Figure \ref{fig:scatter_mean_period}). The RMSD is 1.55 s and 1.86 s, for the mean periods Tm01 and Tm02, respectively. The correlation coefficient is 0.81 for both mean periods. The Doppler correction lowers the RMSD with 0.8 s for both mean periods while the correlation coefficient remains similar (Figure \ref{fig:scatter_tf_ds} and \ref{fig:scatter_tzc_ds}). The original mean periods had a positive MD, which is compensated with the Doppler corrections, especially for Tm01 (Figure \ref{fig:scatter_tf_ds}). 

\begin{figure*}[ht]
\centering
    \begin{subfigure}{0.45\textwidth}
        \includegraphics[width=1\textwidth]{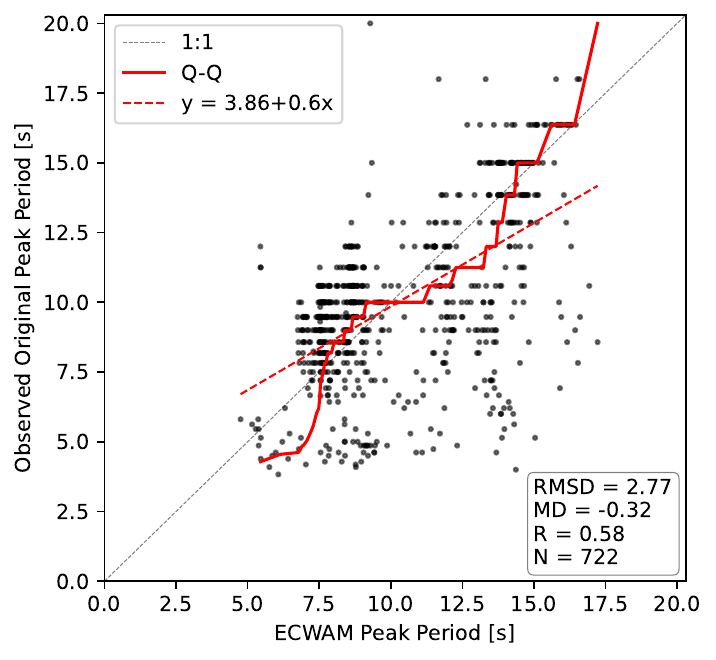}
        \caption{}
	  \label{fig:scatter_tp}
    \end{subfigure}
    \begin{subfigure}{0.45\textwidth}
        \includegraphics[width=1\textwidth]{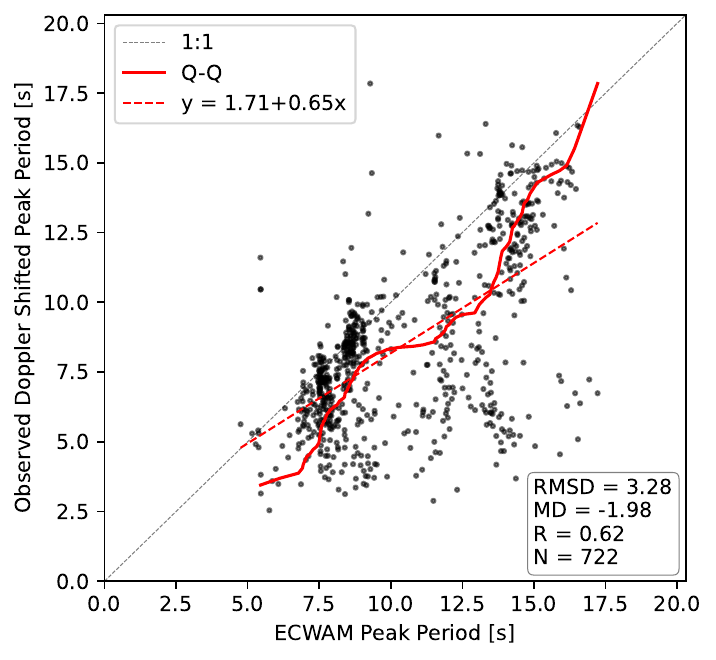}
        \caption{}
	  \label{fig:scatter_tp_ds}
    \end{subfigure}
\caption{Observed (a) original peak period and (b) Doppler shifted peak period, compared to ECWAM's peak period. The solid red line is the quantile-quantile fit (Q-Q), while the red dotted line is the linear regression (y). The statistics are described in Appendix \ref{sec:validation_statistics}.}
\label{fig:scatter_peak}
\end{figure*}

The observed peak period has a mean value of 10.2 s and is limited within the range of 2-20 s by the frequency range 0.05-0.5 Hz. Figure \ref{fig:scatter_tp} displays a spread between the observed peak period and the model data. The RMSD is 2.77 s and the correlation coefficient is 0.58. Figure \ref{fig:scatter_tp} also illustrates that the observed peak period is sorted in steps, with larger spacing at higher periods. This is due to the calculation method (Equation \ref{eq:Tp}) because the division with small frequencies has a larger spacing (1/0.1 Hz = 10s and 1/0.2 Hz = 5 s) than the division with high frequencies (1/0.4 Hz = 2.5 s and 1/0.5 Hz = 2 s). 

The Doppler shifted peak period is not sorted in steps (Figure \ref{fig:scatter_tp_ds}) because the Doppler shift introduces a disturbance dependent on the frequency (Equation \ref{eq:fs_neg}). The Doppler shifted peak period has a 0.51 s higher RMSD than the original peak period. The negative MD is 1.66 s larger in magnitude than the original peak periods which means that the Doppler corrected peak periods are shifted too low compared to the model data. 

The Doppler Effect correction did not improve the peak period measurements as much as a previous study utilizing ship mounted laser \citep{Collins2016}. The study found the statistics to be better when using the wave directions from a marine X-band radar compared to a constant wave direction which is used in the present study. \cite{Collins2016} also found that a method including varying ship speed and heading improved the statistics marginally, compared to constant ship speed and heading which is used in the present study. Moreover, the present study utilized a sailing vessel as a research vessel, which mainly sails with the wind and waves from behind. Steaming with waves results in a reduced Doppler shifted spectral range (Figure \ref{fig:frequency_critical}), which may distort the Doppler shifted peak period. 

\section{Conclusion}\label{sec:conclusion}

This study has presented a ship borne wave measurement system's design and signal processing. The system incorporates a downward facing ultrasonic altimeter with an inertial measurement unit to estimate the sea surface fluctuation as a time series. The system design and software have been optimized during previous studies \citep{Knoblauch2022, Olberg2023}. The best configuration was found to be an ultrasonic altimeter probe with a 15.2 m range, secured within a steel frame at the ship's bow which is elevated 7 m above the sea surface. The inertial measurement unit should be protected from vibrations and physical harm. The 4.5 m horizontal and 1 m vertical distance between the IMU and altimeter probe is sufficient to resemble the same motion. The sea state in the Tropical Atlantic was relatively calm, with a maximum significant wave height of 3.3 m. The system has proven robust when operating over six months, even in rough sea conditions \citep{Olberg2023}. 

The one-dimensional wave spectrum and wave properties have been calculated from the sea surface fluctuation time series. The significant wave height observations fit well with satellite altimetry and the spectral wave model ECWAM. The observed mean periods also agree well with the spectral wave model, while the observed peak period has a lower accuracy. The ship mounted approach is therefore suitable to study the air-sea interaction processes that are more sensitive to the short wind sea than the swell. The Doppler correction enhances the accuracy of the mean periods, but not the peak period. The Doppler correction is anticipated to perform better by utilizing a marine X-band radar for accurate wave direction observations \citep{Lund2017} and for a ship steaming into waves rather than with waves. The potential for measuring directional spectra using multi-sensor setups on ship should also be investigated, as already applied on aircraft \citep{Sun2005, Pettersson2003}.

The findings of this study demonstrate that the integrated wave parameters from ship mounted ultrasonic altimeter have high accuracy. The observations are applicable to real-time broadcasting and validation of satellite observations and wave models. The ship-mounted system has been optimized during the expedition and previous studies \citep{Knoblauch2022, Olberg2023}, to qualify as a robust and adaptable method. The system is based on open source hardware, firmware, and postprocessing, allowing replication of the setup presented here. We believe a culture of sharing off-the-shelf sensors within the scientific community will greatly enhance the volume of in situ observations.

\section*{Acknowledgement}

This work is a part of the One Ocean Expedition, which involves Stiftelsen Seilskipet Statsraad Lehmkuhl and several scientific institutions: The Norwegian Institute of Marine Research, The University of Bergen, The Norwegian Research Center, The Norwegian Meteorological Institute, Western Norway University of Applied Science, Nansen Environmental and Remote Sensing Center, and The Norwegian Institute for Water Research. We would also in particular thank the crew on Statsraad Lehmkuhl for all their practical support to our field work.

\appendix

\section{Open source system design and code}\label{sec:opem_source_code}

All the design information, code, and postprocessing scripts needed to build a similar system and to analyze the data are made available on GitHub at: \url{https://github.com/jerabaul29/OneOcean_wave_ultrasonic_sensor_system} .

This project has highlighted several hardware and software design aspects that are worth keeping in mind and using as guidelines for the future development of similar systems. In particular, it has proven critical to implement the following set of guidelines:

\begin{itemize}
    \item use connectors that are easy and robust to plug / unplug on all sensors; using wires that are permanently fixed makes it hard to draw cables and install the equipment,
    \item protect electronics, in particular the ADC of the microcontroller board, against transient high currents when using a 4-20mA current loop; failure to do so can result in burning ADC due to high transients and parasitic capacity / inductance on transmission lines; this can be avoided either ideally by including schottky diodes to protect the ADC channels, or as a poor mans fix by having strict routines and procedures for plugging / unplugging the instruments, including making sure all instruments are completely off and possible capacitors are fully discharged before any plugging / unplugging sensors,
    \item using a modular design to make it easy to replace components independently of each other as they may experience failure in real world harsh conditions
    \item perform all logging from a single microcontroller board to avoid unreliable timing introduced by non-real-time OS, and clock synchronization issues; i.e., all logging should be performed on the same microcontroller board, rather than merging the data acquired by several microcontroller boards, and / or using the RaspberryPi or similar computer running a non-real-time-operating-system to log some of the sensors,
    \item keep IMU and range sensors as close as possible to each other to simplify the position mismatch-induced rigid body correction
    \item use robust waterproof and shockproof enclosures, and protecting connectors and cables
    \item implementing systematic use of hardware watchdog to avoid freeze of the system in case of unexpected failure, and log all events to allow easy debugging,
    \item for cold regions, include de-icing system on the range sensors to avoid the UG or radar probe to end up covered in ice,
    \item regarding the choice of the VN100 vs. ISM330DHCX-based home assembled IMU: we observe the same intrinsic performance with both systems, so it is possible to drastically cut cost by using the ISM330DHCX-based IMU logging, and fitting it in a proper enclosure,
    \item regarding the choice of the UG vs radar range measurement device: we found out that the UG is both significantly better and cheaper than the radar, and much better suited for use from ships with metal hull,
    \item it is critical to choose good mounting locations on the ship, in particular for the range sensors: the sensors should be mounted so as not to be too exposed to the elements, but at a location where waves are not overly disturbed by the ships hull.
\end{itemize}

\section{Schematic Circuit Diagram}\label{sec:circuit_diagram}

\begin{figure*}[ht]
	\centering
	\includegraphics[width=1\textwidth]{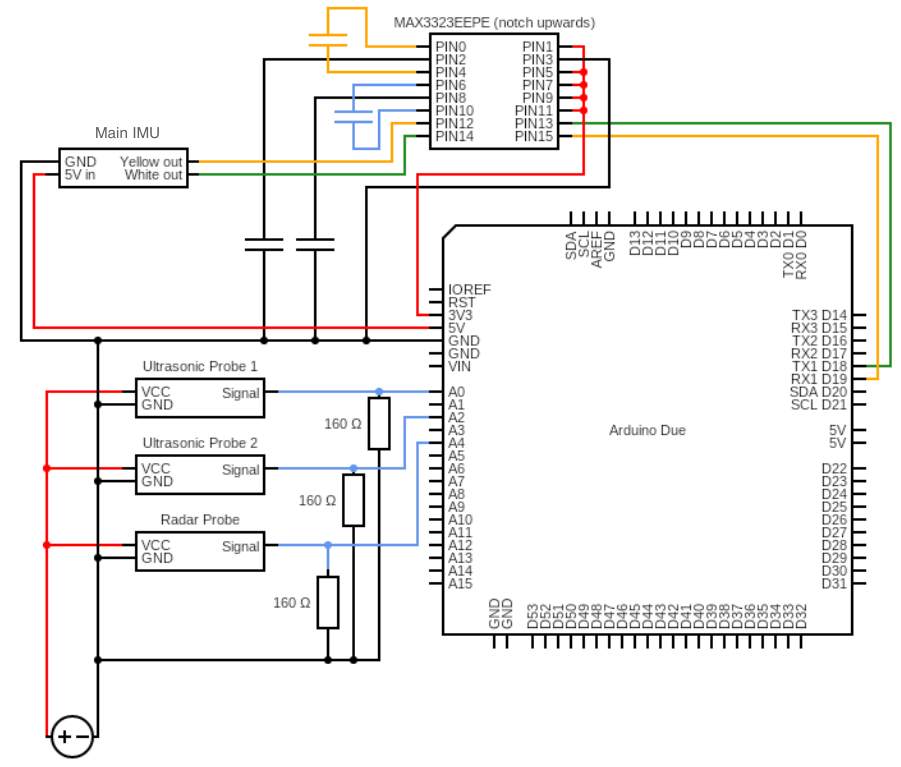}
	\caption{Schematic circuit diagram of the connections between the altimeter probes, the main IMU (VN100), the circuit shield (MAX3323EEPE), and the Arduino Due \citep{Knoblauch2022}. The altimeter probes are supplied with power from a 24V converter plugged into the ship's 220V network. The IMU gets power from the Arduino Due, which receives power through its USB connection with the RPI, providing 5V input (not presented here).}
	\label{fig:circuit_diagram}
\end{figure*}

The electrical diagram of the system is presented in Figure \ref{fig:circuit_diagram}.

\section{Validation Statistics}\label{sec:validation_statistics}

For the validation statistics, the variables are reversed compared to the Wavy documentation\footnote{\url{https://wavyopen.readthedocs.io/en/latest/tutorials_stats.html}}. The wave model and satellite data are treated as the "ground truth" instead of the observations since the purpose is to quantify the measurement system. 

\begin{itemize}
    \item $x_{_o}$ is the observed values of the chosen parameter
    \item $x_{_v}$ is the value to evaluate against as the "ground truth" (satellite and model) 
    \item $N$ is the number of collocated values to be used for verification
\end{itemize}

Bias or mean deviation:
\begin{equation}\label{eq:MD}
    \textrm{\footnotesize{Bias}} = \textrm{\footnotesize{MD}} = \frac{1}{N}\sum_{i=1}^{N}(x_{_o}-x_{_v})_{_i}.
\end{equation}

Root mean squared deviation:
\begin{equation}\label{eq:RMSD}
    \textrm{\footnotesize{RMSD}} = \sqrt{\frac{1}{N}\sum_{i=1}^{N}(x_{_o}-x_{_v})^{2}_{_i}}.
\end{equation} 

Pearson product-moment correlation coefficient:
\begin{equation}\label{eq:corr}
    R_{_{o,v}} =\frac{cov(x_{_o},x_{_v})}{std(x_{_o})\cdot std(x_{_v})}.
\end{equation}



\printcredits

\bibliographystyle{cas-model2-names.bst}

\bibliography{library}

\end{document}